\documentstyle[sprocl,epsfig]{article}

\def\be{\begin{equation}}
\def\ee{\end{equation}}
\def\bea{\begin{eqnarray}}
\def\eea{\end{eqnarray}}
\def\fxv{$f_X/f_V$}

\def\cgs{erg cm$^{-2}$ s$^{-1}$}

\begin{document}

\title{AGNS AMONG ROSAT BLANK FIELD SOURCES}

\author{I. CAGNONI$^{1}$,M. ELVIS$^2$, D.-W. KIM$^2$, J.-S. HUANG$^2$, A. CELOTTI$^3$  and F. NICASTRO$^2$}

\address{$^1$ Dipartimento di Scienze, Universit\`a dell'Insubria, Via Valleggio 11, 22100, Como, Italy (E-mail:ilaria.cagnoni@uninsubria.it)\\$2$ Harvard-Smithsonian Center for Astrophysics, 60 Garden St., Cambridge, MA 02138, USA\\$3$ SISSA-ISAS, Via Beirut 4, 34014, Trieste, Italy}

\maketitle\abstracts{We selected a sample of sources with 
high X-ray over optical flux ratio  from the {\em ROSAT} PSPC archives.
Out of the 16 sources in our sample, one is a (possibly extreme)
BL Lac and two are AGNs.
Six more sources are tentatively classified 
as  AGNs on the basis of their X-ray properties and of our optical-IR imaging. 
For one of the AGNs and for the AGN-like blanks
 we find a discrepancy between the lack of 
absorption indicated by the X-ray  spectra and the presence of
 red counterparts in their error circles.
We discuss various possibilities to reconcile the X-ray and optical information.}

\section{{\em ROSAT} blank field sources}\label{sec_blanks}

We have found a class of `blank field sources' or `blanks' (Cagnoni
et al. 2002). These are X-ray sources with \fxv $>10$ and so are
extremely X-ray loud. We selected our sample from the bright ($f_X >10^{-13}$ erg cm$^{-2}$ s$^{-1}$) high 
latitude ($|b| > 20$) {\em ROSAT} sources in WGACAT (1995 version, 
White Giommi \& Angelini 1984) with no optical
optical counterpart on the POSS (McMahon \& Irwin 1992). 
While our sample contains
only 16 blanks many more may exist, given our demanding search criteria
(e.g. no optical counterpart within 39", 99\% radius). Blanks are a 
varied population. We have previously shown (Cagnoni et al. 2001) that
this is an efficient method of finding high redshift ($z>0.6$) clusters of
galaxies. Here we highlight the peculiar AGNs among the blanks.
(We will use the term AGN in the text meaning unbeamed AGNs,
 i.e. excluding the blazar class).

\section{AGNs expected among {\em ROSAT} blanks}\label{sec_exp}
Some  ``normal'' BL Lacs are expected to contaminate our sample  for $f_X/f_V<35$ (e.g. Maccacaro et al. 1988),
 but blanks reach  $f_X/f_V$ up to a factor of 10 higher.
An obvious possibility is that we are selecting the most extreme representatives of the BL Lacertae class, i.e. extremely variable (a factor of $\geq 10$)
and/or BL Lacs with SED peaking at energies
higher than the {\em ROSAT} band.
This peak could be either the low energy (synchrotron, 
and some of these objects, which also proved to be extremely variable,
 were recently found by Costamante et al. 2001)
or the high energy  (inverse Compton, no such BL Lacs have been found so far) 
one shifted appropriately.\\
\indent Normal type 1 AGNs with $f_X>10^{-13}$ \cgs \ have $O \sim 18$ 
and could be selected as blanks only if their optical counterpart 
is obscured in the O-band by $A_O>3.5$~mag, which corresponds to 
a local $N_H>4 \times 10^{21}$~cm$^{-2}$, or if their optical emission
is intrinsically highly suppressed.
Not all obscured AGNs
could still be visible with $f_X>10^{-13}$ \cgs \/ in the  {\em ROSAT} PSPC.
AGNs can reach such X-ray fluxes despite the obscuring material 
only if they are intrinsically bright, as quasars or QSOs,
or if their  spectrum is dominated by a  conspicuous
soft X-ray excess, as Narrow Line Seyfert galaxies (NLSy).
QSO-2s and NLSy2 galaxies are the ``missing links''
of the unifying models of AGN (Urry \& Padovani 1995) and 
 their number density and importance for the X-ray background 
are still debated (e.g. Comastri et al. 2001).
The other possibility for AGNs to be selected as blanks is when their 
optical emission, which is dominated by the big blue bump (BBB) 
component from the accretion disk (Elvis et al. 1994), is highly suppressed.
This can happen in AGNs powered by  radiatively inefficient
 mechanisms for  accretion, e.g.  advection dominated accretion flows
(ADAFs, e.g. Narayan \& Yi, 1994), where 
most of the energy is stored in the gas and advected toward 
the hole and only a small fraction is radiated.

\section{AGNs and AGN candidates found among blanks}\label{sec_found}

The main observational properties of the AGN and AGN-candidates 
in our sample are summarized in Table~1.
While we were progressing our investigation 
one of the blanks, 1WGA~J1340.1$+$2743,  
was classified as a BL Lac for the lack of features in its optical spectrum 
(Lamer, Brunner \& Staubert (1997).
The source is one of the most extreme in our sample and both its \fxv \/ ($\sim 250$)
 and its broad-band radio indices ($\alpha _{ro} = 0.62$ and $\alpha
_{ox} = 0.61$ for a NVSS counterpart of 4.6~mJy) are not compatible with normal BL Lacs.
1WGA~J1340.1$+$2743 properties can be explained if this source is an extreme 
synchrotron dominated BL Lac such as those in Costamante et al. (2001) 
or if the source  was  flaring  during the WGACAT observation. 
{\em Chandra} follow-up caught the source 10 times fainter than in WGACAT, 
suggesting that the source could indeed be an extreme and/or 
extremely variable BL Lac.\\
\indent Two blanks
(1WGA~J1412.3$+$4355 and 1WGA~J1415.2$+$4403) 
were spectroscopically identified in the RIXOS survey (Mason et al. 2000)
as AGNs at $z\sim 0.5-0.6$. 
1WGA~J1415.2$+$4403 could be a   normal type~1 AGN
because its WGACAT flux was $\sim 50$\% overestimated due to the steep
nature of it spectrum. 
1WGA~J1412.3$+$4355 
presents an  unexpected red   POSS counterpart 
while looks only moderately red in our optical-IR follow-up.
Another 6 sources (see Tab.~1) are likely to be AGNs.
1WGA~J0951.4$+$3916  shows signs of absorption in the X-ray band 
($N_H \sim 2^{+3}_{-1} \times 10^{21}$ cm$^{-2}$) 
 and has a red counterpart (POSS and our imaging).
One more source, 1WGA~J1420.0$+$0625, differs from the AGN-like ones
for the lack of a red counterpart. Could we be seeing the reflected component  of an intrinsically bright obscured AGN?

\subsection{Perverse AGNs: Dusty, yet Unobscured in X-rays}
The remaining  5 AGN-like sources  seem to share the same peculiar combination
of characteristics: 
type~1 AGNs  X-ray spectra  with low absorbing column
 densities (consistent with the Galactic values) but 
red  counterparts ($O-E>2$ in  the POSS and $R-K>4$ in our 
imaging)\footnote{With the exception of  1WGA~J1535.0$+$2336, 
red in the POSS, but only moderately red in our imaging;
of  1WGA~J1220.6$+$3347 red in our imaging but not in the POSS}.
For a local AGN type~1, $O-E>2$ implies $A_V > 4$ (e.g. Risaliti et al. 2001) 
and thus strong obscuration ($N_H > 2 \times 10^{21}$ cm$^{-2}$ 
for a dust to gas ratio typical of the ISM in our Galaxy),
in contrast to the results of the X-ray data.\\
There are some possibilities to reconcile the optical and X-ray data:
(1) the AGNs have $z\geq 3.5$ (e.g. Fig.~7 of Risaliti et al. 2001). 
This is excluded for 1WGA~1415.2$+$4403, 
but it is still possible for the other RIXOS AGN (1WGA~J1412.3$+$4355)
 whose the redshift measurement is based on only 1 optical line;
(2) a dust to gas ratio $\sim 40-60$ times higher than the Galactic value;
(3) a modified  dust grain size distribution  with respect to the Galactic ISM;
(4) the presence of dust within a  warm absorber (e.g. Komossa \& Bade 1998);
(5) a highly suppressed BBB emission (see \S ~\ref{sec_exp}).\\
The planned optical spectroscopic follow-up is needed to
investigate the nature of the AGN-like blanks.\\

\begin{table} 
\footnotesize
\caption{Properties of the AGN blanks.}
\begin{center} 
\begin{tabular}{c c c c c c c c  c c } \hline
Source name     &Identification &$\Gamma$ 	&$F_X$	 &R   &E$^a$   &O$^a$  &K      &O-E$^b$ &R-K\\
(1WGA~J)        &		&	&(.3-3.5 keV)\\
\hline
1340.1$+$2743    &BL Lac        &2.35       &8.94	 &21.3 &-        &-      &$15.8 \pm 1.2$  	&-              &5.5\\
\hline
1412.3$+$4355    &AGN(z=0.59)   &1.76       &0.97	&19.9 &18.73    &-      &$16.7 \pm 0.4$         &$>2.77$        &3.2\\
1415.2$+$4403    &AGN(z=0.56)   &2.97       &0.64	&20.7 &-        &-      &$16.0 \pm 0.4$         &-              &4.7\\
0951.4$+$3916    &AGN?          &2.62       &1.39	&20.1 &19.1     &-      &$16.2 \pm 0.4$         &$>2.4$         &3.9\\
1220.3$+$0641    &AGN?          &1.95       &8.22	&17.6 &18.58    &21.81  &$13.6 \pm 0.4$         &3.23           &4.0\\
1416.2$+$1136    &AGN?          &2.98       &0.59	&19.9 &19.5     &22.22  &$14.9 \pm 0.4$         &2.72           &5.0\\
1535.0$+$2336    &AGN?          &0.70       &0.83	&19.9 &19.6     &21.8   &$16.5 \pm 0.4$         &2.2            &3.4\\
1233.3$+$6910    &AGN?          &2.43       &4.11	&$\sim 22$ &-   &-      &$16.1 \pm 1.2$         &-              &5.9\\
1220.6$+$3347    &AGN?          &2.31       &1.30	&19.9    &20.9  &21.7   &$15.6 \pm 0.6$         &0.8            &4.3\\
\hline
1420.0$+$0625    &Unknown       &2.21       &1.68	&20.0 &-        &-      &$16.6 \pm 0.4$         &-              &3.4\\
\hline
\end{tabular}
\end{center}
\end{table}

\section*{Acknowledgments}
This work was supported by NASA ADP grant NAG5-9206 
and by the Italian MIUR (IC and AC). IC acknowledges a CNAA fellowship.

\section*{References}

\end{document}